\documentclass[lettersize,journal]{IEEEtran}
\usepackage{amsmath,amsfonts}
\usepackage{algorithmic}
\usepackage{algorithm}
\usepackage{array}
\usepackage[caption=false,font=normalsize,labelfont=sf,textfont=sf]{subfig}
\usepackage{textcomp}
\usepackage{stfloats}
\usepackage{url}
\usepackage{verbatim}
\usepackage{graphicx}
\usepackage{multirow}
\usepackage{bm}

\usepackage[numbers]{natbib}
\usepackage{cleveref}

\begin{document}

\title{Ultrasonic Image's Annotation Removal: A Self-supervised Noise2Noise Approach}

\author{Yuanheng Zhang,
    Nan Jiang,
    Zhaoheng Xie,
    Junying Cao*,
    Yueyang Teng*
    \thanks{Y. Zhang is with the College of Medicine and Biological Information Engineering, Northeastern University, China.}
    \thanks{N. Jiang is with the Department of Ultrasound, General Hospital of Northern Theater Command, China.}
    \thanks{Z. Xie is with the Institute of Medical Technology, Peking University, China.}
    \thanks{J. Cao is with the Department of Ultrasound, General Hospital of Northern Theater Command, China.}
    \thanks{Y. Teng is with the College of Medicine and Biological Information Engineering, Northeastern University, China.}
    \thanks{J. Cao and Y. Teng contributed equally to this work.}
    \thanks{This work is supported by the Natural Science Foundation of Liaoning Province (2022-MS-114).}
    \thanks{This work is supported by the Key R\&D Plan Projects of Liaoning Province in 2020 (Project No. 2020JH2/10300122).}
    }

\markboth{IEEE Transactions on Computational Imaging}
{Zhang, Jiang\MakeLowercase{\textit{et al.}}: Ultrasonic Image's Body Marker Annotation Removal: A Noise2Noise Approach}

\maketitle

\begin{abstract}
    Accurately annotated ultrasonic images are vital components of a high-quality medical report.
    Hospitals often have strict guidelines on the types of annotations that should appear on imaging results.
    However, manually inspecting these images can be a cumbersome task.
    While a neural network could potentially automate the process, training such a model typically requires a dataset of paired input and target images, which in turn involves significant human labour.
    This study introduces an automated approach for detecting annotations in images.
    This is achieved by treating the annotations as noise, creating a self-supervised pretext task and using a model trained under the Noise2Noise scheme to restore the image to a clean state.
    We tested a variety of model structures on the denoising task against different types of annotation, including body marker annotation, radial line annotation, etc.
    Our results demonstrate that most models trained under the Noise2Noise scheme outperformed their counterparts trained with noisy-clean data pairs.
    The costumed U-Net yielded the most optimal outcome on the body marker annotation dataset, with high scores on segmentation precision and reconstruction similarity.
    We released our code at \url{https://github.com/GrandArth/UltrasonicImage-N2N-Approach}.
\end{abstract}


\section{Introduction}\label{sec: intro}

\IEEEPARstart{A}{nnotations}, typically comprised of various labels and marks, are commonly utilized to record critical information from an ultrasonic exam, 
including the precise location of potential lesions or suspicious findings, on archived results. 
Such annotations prove beneficial in aiding physicians in interpreting the exam results, 
particularly when surrounding structures do not provide any indication of the anatomic location of the image. 
Additionally, hospitals often mandate the inclusion of annotations, especially in cases involving inter-hospital patient transfers \cite{kulshrestha2016inter}. 
If the report does not have comprehensive annotations, patients are usually required to undergo an equivalent radiography exam at the facility of transfer. 

Commonly employed types of annotations include body marker annotation \cite{jackson2017ultrasonic}, radial line annotation, and vascular flow annotation.
The presence of these annotations serves as evidence for the standardization of the diagnostic process. Annotations not only document the reasoning behind the diagnostic assessment but also facilitate comparison between pre- and post-treatment imaging findings to gain further insight into the patient's condition. 

However, the utilization of annotations during ultrasound exams may vary depending on the proficiency of the sonographer performing the procedure. 
Ultrasound being a live examination makes it hard to implement additional reviews, thereby relying solely on the expertise of the operator to determine the presence of annotations. 
Furthermore, the need for repetitive manual verification increases the likelihood of forgetting the task, particularly during busy schedules at hospitals. 
As such, it is possible for the absence of annotations to occur.

Given the strict regulations and obvious beneficiation surrounding the need for annotations in medical imaging, 
sonographers need to manually validate that the stored data satisfies these requirements to ensure that diagnoses meet the standard continuously.
However, this is a cognitively demanding undertaking as it entails the fulfillment of diverse annotation obligations tailored to specific image outcomes. 
In addition, dealing with archived files manually is a cumbersome task as most medical data management systems do not consider this necessary and have no relevant feature implemented.

The utilization of neural networks for the automatic assessment of whether the stored data meets particular criteria is a logical approach. 
To address the current issue, there are several approaches that can be taken using different types of deep learning models. 
The first approach would involve treating the task as a semantic segmentation problem, where the goal is to classify each pixel in the image into one of several predefined categories. 
Alternatively, the task could be framed as an instance segmentation problem, where the aim is to identify and label individual objects within the scene. 
In order to accomplish these goals, attention-based models such as the Pyramid Attention Network \cite{li2018pyramid} or the Reverse Attention Network \cite{huang2017semantic} could be employed. Alternatively, generative models like variants of Generative Adversarial Networks (GANs) \cite{goodfellow2020generative} are also viable. 
Once the segmentation has been completed, the resulting labels could then be used to determine whether the image meets regulatory requirements or not. 
This task could also be viewed as an object recognition challenge, and for this purpose, 
models such as Single Shot MultiBox Detector (SSD) \cite{liu2016ssd} or You Only Look Once (YOLO) \cite{redmon2016you} could be utilized to obtain the four coordinates of the bonding box of a detected object, which will serve as demonstrative evidence of the necessary annotations.
In order to train a model using deep learning, it is important to have a suitable training dataset that includes paired input and output data, regardless of the specific task being performed.
However, building an appropriate training dataset is a challenging task due to the absence of high-quality data such as segmentation masks, object coordinates and clean targets. 
Acquiring such data requires a considerable amount of manual effort. 

In this study, we introduced a self-supervised Noise2Noise approach to recognise annotations without needing a pairwise dataset by manually superposing common annotations onto a small set of unannotated images randomly and repeatedly. 
We trained multiple network structures such as FCN, U-Net++, MultiResUNet, etc., for Noise2Noise to select an ideal one.
We noted that the majority of Noise2Noise based methods surpassed the corresponding
Noise2Clean (supervised learning) methods in which the former even receive a Sørensen-Dice coefficient (Dice) increases of up to $300\%$, an Intersection over Union (IoU) increase of up to $384\%$, and a Peak Signal to Noise Ratio Human Visual System Modified (PSNR\_HVS\_M) increase of up to $38\%$ in some cases.
Among them our costumed U-Net achieved the best results, both quantitative and qualitatively.

The remainder of the paper is organized as follows:
Section \ref{sec: background} discusses related works.
Section \ref{sec: methods} outlines our methodology, data sources, dataset building pipeline and model strcutures used in this work.
In Section \ref{sec: results},
quantitative metric scores and qualitative image results are provided to support our claim regarding the optimal model structure, loss function and observations on Noise2Noise's effect.
Finally, Section \ref{sec: discuss} concludes the paper.

\section{Related Works} \label{sec: background}

\subsection{Self-supervised Learning}\label{sec:ssl}

Self-supervised learning is a way of training deep-learning models without human guidance or explicit instructions. 
Unlike supervised learning which uses labeled examples, self-supervised models learn from unlabeled data by identifying patterns and relationships on their own.
It uses the structure of images (e.g., edges, shapes) to teach the deep-learning model how to identify important parts of an image automatically, rather than having to be explicitly told what to look for.
This is particularly helpful considering the abundance of unlabeled data that exists today and the amount of work required to create a properly constructed dataset.
To create a robust, large model, self-supervised learning is an essential tool.

The general process of self-supervised learning involves first creating a pretext task for the model to solve. By completing this task, the model can gain an understanding of the structural information embedded within the data. This understanding can then be transferred to downstream tasks using different forms of transfer learning. 

Examples of pretext tasks include rotating an image for the model to predict the degree of rotation, reconstructing images from an altered view, or reconstructing images from a corrupted version of the original data.

In this work, we developed a pretext task where we asked the model to generate another noisy image from the noisy input while keeping the same original clean image beneath it. 
Specifically, we manually extracted several common annotations from stored data and randomly superimposed them on a small set of unannotated images to create a large dataset.
The idea behind this approach was to train the model to recognize the crucial features of the original so that it could distinguish between noise and clean images.

\subsection{Noise2Noise Training Scheme}\label{sec:n2nback}

Noise2Noise is originally proposed in  \cite{lehtinen2018Noise2Noise} as a novel statistical reasoning for the task of image denoising.
It is shown that, under certain key constraints, 
it is possible to train a denoising model using only corrupted images. 
The constraints are: the distribution of the added noise must have a mean of zero and no correlation with the desired clean image, 
and the correlation between the noise in the input image and the target image should be close to zero \cite{kashyap2021speech}. 

By utilizing deep learning, a denoising task can be transformed into a regression problem, 
where a neural network is used to learn the mapping between corrupted samples $\hat{x}_i$ and clean samples $y_i$ by minimizing the empirical risk \cite{lehtinen2018Noise2Noise}

In  \cite{lehtinen2018Noise2Noise}, inspecting the form of a typical training process shows that training a neural network is a generalization of a point estimating problem.
We can see that it is essentially solving the point estimating problem for each separate input. 
This means, by finding the optimal parameters, the trained neural network will output the expectation or median of all possible mapping for input $x$. 
This property often leads to unwanted fuzziness in many deep-learning applications.
However, in a denoising scenario, when the noise satisfies the above constraints and exists in both the model input and training target, the task of empirical risk minimization, given infinite data, 

\begin{equation}\label{eq4}
    \underset{\theta}{\operatorname{argmin}} \sum_i L\left(f_\theta\left(\hat{x}_i\right), \hat{y}_i\right)
\end{equation}

is equivalent to the original regression problem

\begin{equation}\label{eq1}
    \underset{\theta}{\operatorname{argmin}} \sum_i L\left(f_\theta\left(\hat{x}_i\right), y_i\right)
\end{equation}

where $f_\theta(x)$ is the model parameterized by $\theta$, $L$ is loss function, $\hat{x}_i,\hat{y}_i$ are samples drawn from a noisy distribution and $y_i$ representing clean samples.

The idea of using the self-supervised learning in conjunction with Noise2Noise training scheme aligns well with our goal of obtaining a clean image. 
With a clean image, we can easily produce a segmentation map for various kinds of annotations, facilitating the models to recognise and categorize them accurately.

\section{Methodology}\label{sec: methods}

Initially, our data includes collections of information that may or may not have specific annotations. 
We manually examined and filtered the data to create a clean dataset for each annotation.
Next, we studied the individual components of different annotations and identified a general pattern for each one. 
Using this pattern, we generated large datasets containing noisy data and trained a denoising model using the Noise2Noise approach.
Finally, we trained various model structures using both the Noise2Noise and conventional Noise2Clean techniques to obtain denoising models for the purpose of performance comparison.

\subsection{Dataset}

To manually synthesize a self-supervised Noise2Noise dataset, which our training requires,
it is essential to know the scheme of the different annotations and to construct a dataset according to it.

Our original data consists mainly of ultrasonic images provided by the General Hospital of Northern Theater Command.
These images were captured using external video capture cards and are in 8-bit sRGB format.

According to the type of noise, we divided these data into six categories:
\begin{itemize}
    \item Images with body marker annotation
    \item Images without body marker annotation
    \item Images with radial line annotation
    \item Images without radial line annotation
    \item Images with vascular flow annotation
    \item Images without vascular flow annotation
\end{itemize}

\begin{figure*}[!t]
    \centering
    \subfloat[]{\includegraphics[height=5cm]{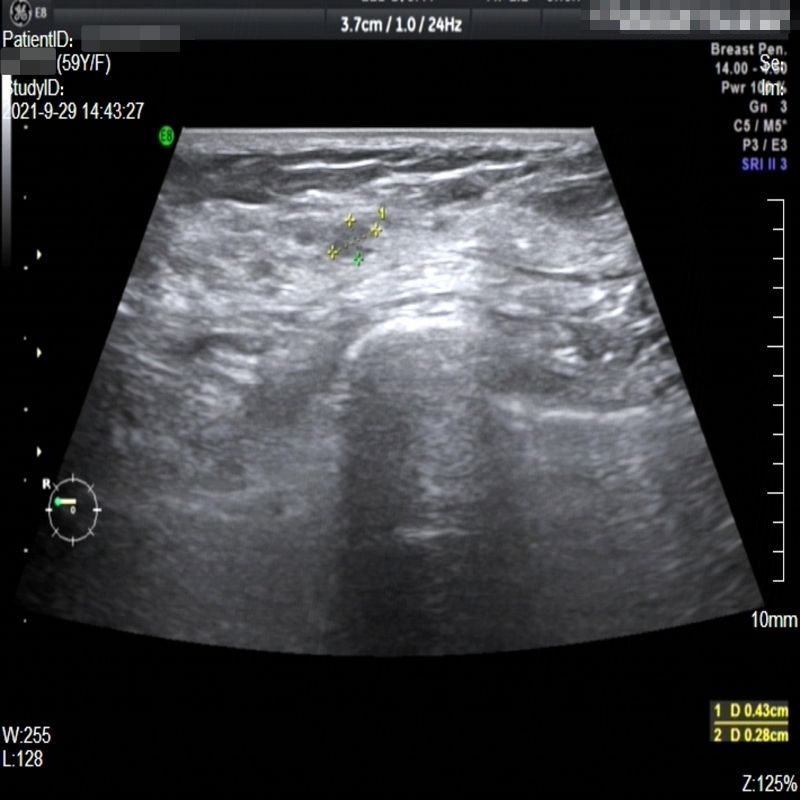}}
    \hfil
    \subfloat[]{\includegraphics[height=5cm]{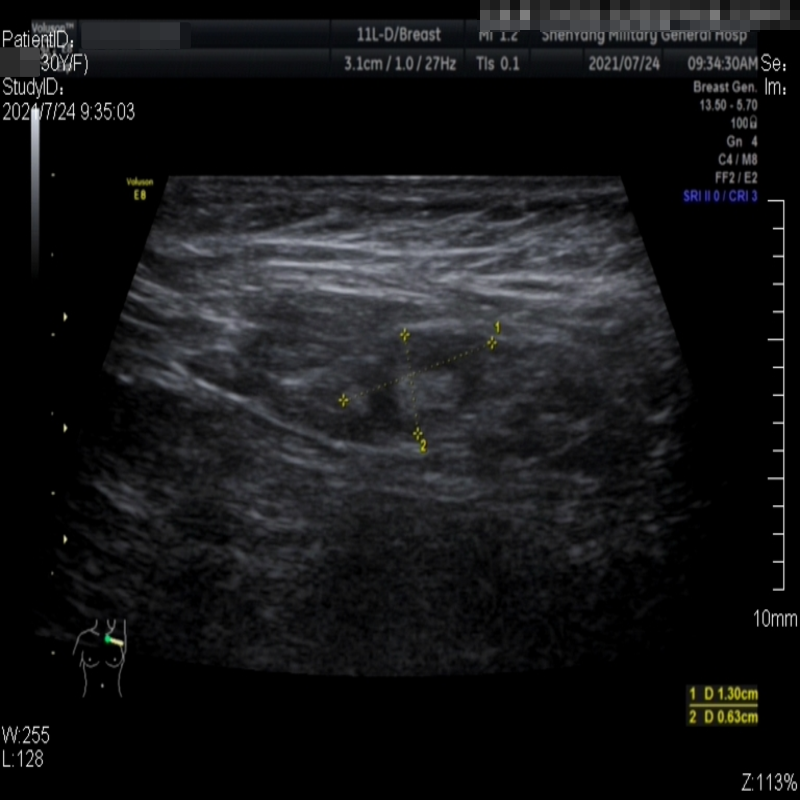}}
    \hfil
    \subfloat[]{\includegraphics[height=5cm]{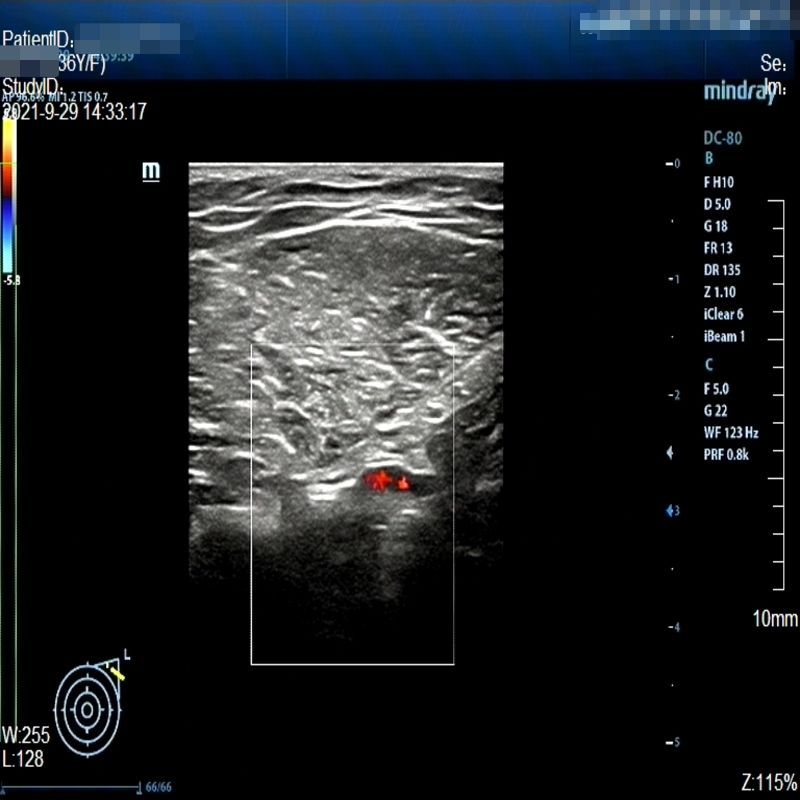}}
    \hfil
    \caption{Images with various annotations. (a) body marker annotation, (b) radical line annotation, (c) vascular flow annotation.}
    \label{fig: Sample Images With Annotation}
\end{figure*}

Images with certain annotations are considered noisy images in the context of the noise removal task, and corresponding images without these annotations are considered clean.
Some typical images with various annotation are provided in Fig. \ref{fig: Sample Images With Annotation}.

To safeguard the confidentiality of the patient, any personal data displayed in the margin of the image is blurred using pixelization. This same technique is also used to obscure any similar information present in other images.


In essence, a body marker annotation is a marker selected from a fixed set of icons that indicates different regions of the human body and its current orientation.
It is typically located at the edge of the ultrasnoic image area and is labeled by the sonographer.
On some ultrasound machines, the body marker annotation has a fixed position.

However, from a statistical and training perspective, each real instance can be viewed as an image sample from a conditional distribution where the condition is the body marker annotation's location.
By randomly placing body marker annotation at any position within the image, we draw samples from a distribution without the aforementioned condition.
By learning to denoise samples from the unconditioned distribution, the model can effectively denoise samples from conditional distribution as well.

Other commonly used annotations that we introduced later comply with the same reasoning.

The radial line annotation is pairs of connected cross markers.
They are usually placed at the edge of the lesion area, with its placement determined by the size of the lesion.
One to three pairs of cross markers may be present in an image, corresponding to the three axes of 3D space, but typically there are only two pairs.

The vascular flow annotation is not an additional labeling feature meant to simplify identification.
Rather, it serves as a bounding box that identifies the specific area of the image being examined by the ultrasound flowmeter.
However, to keep things simple, we will continue to call it a form of annotation.
The presence of this annotation indicates that the relevant examination has been conducted.

To synthesize a Noise2Noise training dataset for above annotations,
we first manually extracted the necessary annotation icons from existing annotated data,
then we randomly overlay different annotations on the clean images we have.
The randomness of the noise overlay allows for the creation of a relatively large dataset.

By constructing training datasets in the above-mentioned process, each noisy image has three corresponding images for different tasks.

\begin{itemize}
    \item A clean image which the noisy image originated from.
    \item A different noisy image created from the same clean image, using a different (in terms of position, form, etc.) noise sampled from the same distribution.
    \item A binary image recorded the position and form of the noise appended to the clean image.
\end{itemize}

\begin{figure*}[!t]
    \includegraphics[width=\textwidth]{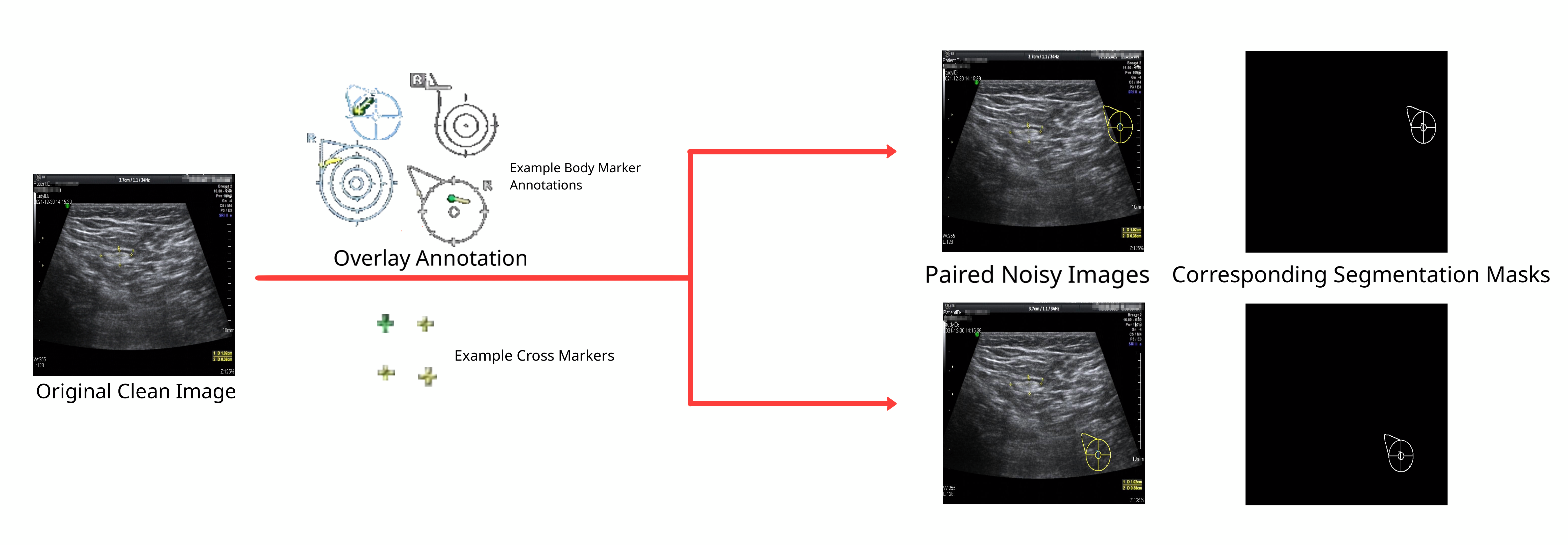}\centering
    \caption{Flow Chart of Training Dataset Building}
    \label{fig: flow chart}
\end{figure*}

An instance of the training dataset is presented in Fig. \ref{fig: flow chart}.
Using these images, the same dataset can be used for Noise2Noise training, conventional Noise2Clean training, and normal segmentation training.

Our approach to create this training dataset can minimize the amount of human labor required. Even with a limited amount of clean data, we are able to generate a large noisy dataset for training. The flow chart of the above process is also shown in Fig. \ref{fig: flow chart}.

\subsection{Network Structures}

In this research, we trained several structures to find the optimal solution and compare the two different training schemes: Noise2Noise and traditional Noise2Clean.

We adopted most of the structures from the traditional image segmentation model.
The models we adopted include FCN, DeepLabv3, LinkNet, MANet, U-Net Plus Plus, MultiResUNet and a costumed U-Net.

FCN is one of the models utilizing convolutional networks in semantic segmentation.
\citeauthor{long2015fully} \cite{long2015fully,minaee2021image} uses fully-convolutional layers instead of fully-connected layers so that this model is compatible with non-fixed sized input and ouputs.

DeepLabv3 is a subsequent model of the DeepLab model famlily, developed by  \citeauthor{chen2017rethinking} \cite{chen2017rethinking}. 
The main feature of this model is the use of dilated convolution, also known as ``atrous'' convolution. 
This method is advocated to combat the issue of feature resolution reduction in deep convolutional networks (due to pooling operations and strides in convolution operations) and the difficulties in multi-scale segmentation. 

LinkNet is proposed by  \citeauthor{chaurasia2017linknet} \cite{chaurasia2017linknet} to address the problem of the long processing time of most segmentation models.
By using a skip connection to pass spatial information directly to the corresponding decoder, LinkNet manages to preserve low-level information without additional parameters and re-learning operations. 

MANet, or Multi-scale Attention Net, is developed to improve accuracy in semantic segmentation of remote sensing images.
By using a novel attention mechanism, treating attention as a kernel function,  \citeauthor{li2021multiattention} \cite{li2021multiattention,Iakubovskii:2019} reduces the complexity of the dot-product attention mechanism to $O(N)$.

U-Net is a well-known encoder-decoder segmentation model.
It is originally proposed by  \citeauthor{ronneberger2015u} \cite{chlap2021review,ronneberger2015u} for segmenting biological microscopy images. 

U-Net++ is a variant of U-Net proposed by \citeauthor{zhou2018unet++} \cite{zhou2018unet++}.
In their work, they proposed a novel skip connection block in which a dense convolution block is used to process the input from the encoder feature map so that the semantic level of the input is closer to the corresponding decoder feature map.  

MultiResUNet is another modern variant of U-Net proposed by \citeauthor{ibtehaz2020multiresunet} \cite{ibtehaz2020multiresunet} as a potential successor.
They used an Inception-like layer to replace the consecutive convolution layers after each pooling and transpose-convolution layers, to percept objects at different scales. 
They adopted a chain of convolution layers with residual connections instead of plain skip connection to process the feature map inputs before concatenating them to decoder feature maps. 

In our work, since the vanilla U-Net does not match the spatial resolution of our dataset, we used a costumed U-Net similar to  \cite{lehtinen2018Noise2Noise} in all of our tests. 
Convolution layers with different stride and padding are used in this structure to ensure the input and output dimension is identical.

\section{Experimental Results}\label{sec: results}

In this section,
we provide quantitative and qualitative results to support our claim in Section \ref{sec: intro}.

\subsection{Evaluation}
 We evaluate the models performance based on segmentation precision and reconstruction similarity.

 \subsubsection{Segmentation Precision}

 In terms of noise reduction precision, for a typical segmentation model, we can use the output to compare it with a binary image known as the truth mask to compute a score based on the number of pixels that get classified into the right categories. 
 For a restoration model like ours, we subtract the model output from the model input to compute the binary segmentation result. 
 We compare the results with the segmentation truth mask to compute the Dice, IoU, and Pixel Accuracy (PA).

\subsubsection{Reconstruction Similarity}

For assessing reconstruction similarity, we use two metrics: Structural Similarity Index Measure (SSIM) and PSNR\_HVS\_M. 
SSIM is a commonly used measure of image similarity.
The PSNR metric known as PSNR\_HVS\_M  \cite{ponomarenko2007betweenpsnrhvsm} is considered to be a more accurate representation of image quality,
which takes into consideration the Contrast Sensitivity Function (CSF) and the between-coceptor contrast masking of Discrete Cosine Transform (DCT) basis functions.

\subsection{Training}
The neural networks discussed in the previous section were trained using PyTorch 1.10.1. 
RMSprop \cite{tieleman2012lecture}, a variant of stochastic gradient descent that divides gradients by an average of their recent magnitude, was used as the optimizer with a learning rate of $0.00001$, momentum of $0.9$, weight decay of $1e-8$, and default values \cite{rmsproptorch} for other parameters.

Three datasets were created in aforementioned process to train various denoising models.
For body marker annotation, a dataset of $83,900$ pairs of noisy images generated from $4,975$ clean images was used. 
For radial line annotation, $80,000$ pairs of noisy images were generated from $3,936$ clean images. 
For vascular flow annotation, $80,000$ pairs of noisy images were generated from $250$ clean images.

\subsection{Optimal Model Structure}\label{sec:cross}

To find the most effective combination of network structure and training scheme for the given task, we trained different network structures under the Noise2Noise and Noise2Clean schemes using the body mark annotation dataset.
Though utilizing only one type of annotation, this experiment's results could demonstrate the likely most suitable structure for other annotations as well.
$L_1$ loss is used to trained these models.
The results were compared using segmentation precision and reconstruction similarity, and are presented in Tables \ref{tab:seg_body_marker_seg} and \ref{tab:seg_body_marker_restore}.

We observed that Noise2Noise training scheme improves segmentation precision and reconstruction similarity in most cases.
The results presented in Tables \ref{tab:seg_body_marker_seg} and \ref{tab:seg_body_marker_restore} indicate that the models trained using the Noise2Noise scheme generally achieved higher Dice scores, IoU scores, PA scores, and PSNR\_HVS\_M scores. 
Specifically, for the costumed U-Net, we observed an increase in the Dice and IoU of $0.151$ and $0.155$, respectively, and an increase of $11.625$ for the PSNR\_HVS\_M when using linearly normalized input.
According to our hypothesis, the Noise2Noise training process improves the model's ability to understand the features of annotations through solving an ``impossible'' task of relocating the annotation. 
This task is essentially a self-supervised pretext training task that helps the model gain a better understanding of the annotations and the spatial structure of the ultrasonic images, thus gaining higher performance.

We also noted that the costumed U-Net structure performed the best out of all the structures tested.
It achieved the highest Dice, IoU, SSIM, and PSNR\_HSV\_M scores under both training schemes. 
The costumed U-Net trained using the Noise2Noise scheme achieved the highest segmentation precision and reconstruction similarity of all models, with a Dice of $0.712$, an IoU of $0.596$, an SSIM of $0.967$, and a PSNR\_HVS\_M of $41.628$.
Given the above results, we chose the costumed U-Net as the optimal model for later experiments.


\begin{table*}[ht]
    \centering
    \caption{Segmentation Precision on Body Marker Annotation (Average + Var)\\N2C stands for Noise2Clean, N2N stands for Noise2Noise\\ SMN indicates the model is trained with data normalized according to standard deviation and mean\\Models without SMN is trained with linearly normalized data
    }
    \begin{tabular}{|c | c c c c|}
        \hline
        Method                           & Training Mode       & Dice                  & IoU                   & PA                              \\ [0.5ex]
        \hline
        \multirow{2}{*}{FCN\_101}        & N2C SMN             & $0.07\pm0.003$        & $0.039\pm0.001$       & $0.97\pm7.2 \times e^{-5}$      \\
        \cline{2-5}
                                         & \textbf{N2N SMN}    & $\bm{0.07 \pm 0.003}$ & $\bm{0.04 \pm 0.001}$ & $\bm{0.97 \pm 8 \times e^{-5}}$ \\
        \hline

        \multirow{2}{*}{DeepLab V3  }    & N2C                 & $0.073\pm0.003$       & $0.039\pm0.001$       & $0.969\pm 0.005$                \\
        \cline{2-5}
                                         & \textbf{N2N }       & $\bm{0.074\pm0.003}$  & $\bm{0.04\pm0.001}$   & $\bm{0.969\pm 0.005}$           \\
        \hline

        \multirow{2}{*}{LinkNet}         & \textbf{N2C}        & $\bm{0.447\pm0.105}$  & $\bm{0.346\pm0.007}$  & $\bm{0.976\pm 0.008}$           \\
        \cline{2-5}
                                         & N2N                 & $0.343\pm0.139$       & $0.280\pm0.106$       & $0.938\pm 0.008$                \\
        \hline

        \multirow{2}{*}{MANet}           & N2C                 & $0.531\pm0.113$       & $0.430\pm0.091$       & $0.943\pm 0.015$                \\
        \cline{2-5}
                                         & \textbf{N2N}        & $\bm{0.543\pm0.128}$  & $\bm{0.451\pm0.105}$  & $\bm{0.917\pm 0.024}$           \\
        \hline

        \multirow{2}{*}{U-Net++}         & N2C                 & $0.551\pm0.08$        & $0.437\pm0.07$        & $0.983\pm 0.007$                \\
        \cline{2-5}
                                         & \textbf{N2N}        & $\bm{0.613\pm0.114}$  & $\bm{0.516\pm0.09}$   & $\bm{0.943\pm 0.016}$           \\
        \hline

        \multirow{2}{*}{MultiResUNet}    & N2C SMN             & $0.416\pm0.05$        & $0.594\pm0.04$        & $0.998\pm 2.75 \times e^{-6}$   \\
        \cline{2-5}
                                         & \textbf{N2N SMN   } & $\bm{0.661 \pm 0.06}$ & $\bm{0.539\pm0.06}$   & $\bm{0.99\pm 5 \times e^{-4}}$  \\
        \hline

        \multirow{4}{*}{Costumed U-Net  } & N2C SMN             & $0.408 \pm 0.05$      & $0.286\pm0.04$        & $0.998 \pm 2.54 \times e^{-6}$  \\
        \cline{2-5}
                                         & \textbf{N2N SMN  }  & $\bm{0.676\pm0.05}$   & $\bm{0.552\pm0.05}$   & $\bm{0.999\pm 5 \times e^{-7}}$ \\
        \cline{2-5}
                                         & N2C                 & $0.561\pm0.077$       & $0.441\pm0.072$       & $0.990\pm 0.005$                \\
        \cline{2-5}
                                         & \textbf{N2N}        & $\bm{0.712\pm0.053}$  & $\bm{0.596\pm0.058}$  & $\bm{0.993\pm 0.007}$           \\
        \hline
    \end{tabular}
    \label{tab:seg_body_marker_seg}
\end{table*}

\begin{table*}[ht]
    \centering
    \caption{Reconstruction Similarity on Body Marker Annotation (Average + Var)}
    \begin{tabular}{| c | c c c|}
        \hline
        Method                           & Training Mode    & SSIM                   & PSNR\_HVS\_M             \\
        \hline
        \multirow{2}{*}{FCN\_101}        & \textbf{N2C}     & $\bm{0.459 \pm 0.001}$ & $\bm{10.264 \pm 1.751}$  \\
        \cline{2-4}
                                         & N2N              & $0.453 \pm 0.016$      & $10.181 \pm 2.430$       \\
        \hline
        \multirow{2}{*}{DeepLab V3  }    & \textbf{N2C}     & $\bm{0.680 \pm 0.004}$ & $\bm{15.919 \pm 2.578}$  \\
        \cline{2-4}
                                         & N2N              & $0.678 \pm 0.005$      & $15.827 \pm 3.282$       \\
        \hline
        \multirow{2}{*}{LinkNet}         & N2C              & $0.933 \pm 0.000$      & $25.691 \pm 6.425$       \\
        \cline{2-4}
                                         & \textbf{N2N}     & $\bm{0.945 \pm 0.000}$ & $\bm{26.307 \pm 8.466}$  \\
        \hline
        \multirow{2}{*}{MANet}           & N2C              & $0.923 \pm 0.002$      & $21.920 \pm 7.015$       \\
        \cline{2-4}
                                         & \textbf{N2N}     & $\bm{0.923 \pm 0.002}$ & $\bm{23.027 \pm 3.903}$  \\
        \hline
        \multirow{2}{*}{U-Net++}         & N2C              & $0.923 \pm 0.000$      & $21.245 \pm 1.846$       \\
        \cline{2-4}
                                         & \textbf{N2N}     & $\bm{0.927 \pm 0.000}$ & $\bm{24.366 \pm 7.121}$  \\
        \hline
        \multirow{2}{*}{MultiResUNet}    & \textbf{N2C SMN} & $\bm{0.856 \pm 0.002}$ & $\bm{23.712 \pm 3.936}$  \\
        \cline{2-4}
                                         & N2N SMN          & $0.792 \pm 0.004$      & $21.256 \pm 6.160$       \\
        \hline
        \multirow{4}{*}{Costumed U-Net  } & N2C SMN          & $0.833 \pm 0.003$      & $11.828 \pm 20.299$      \\
        \cline{2-4}
                                         & N2N SMN          & $0.791 \pm 0.004$      & $20.746 \pm 10.223$      \\
        \cline{2-4}
                                         & N2C              & $0.961 \pm 0.000$      & $29.976 \pm 30.140$      \\
        \cline{2-4}
                                         & \textbf{N2N}     & $\bm{0.967 \pm 0.000}$ & $\bm{41.628 \pm 41.775}$ \\
        \hline
    \end{tabular}
    \label{tab:seg_body_marker_restore}
\end{table*}


\subsection{Optimal Loss Function}\label{sec:loss}

\begin{figure*}[!t]
    \centering
    \subfloat[]{\includegraphics[height=5cm]{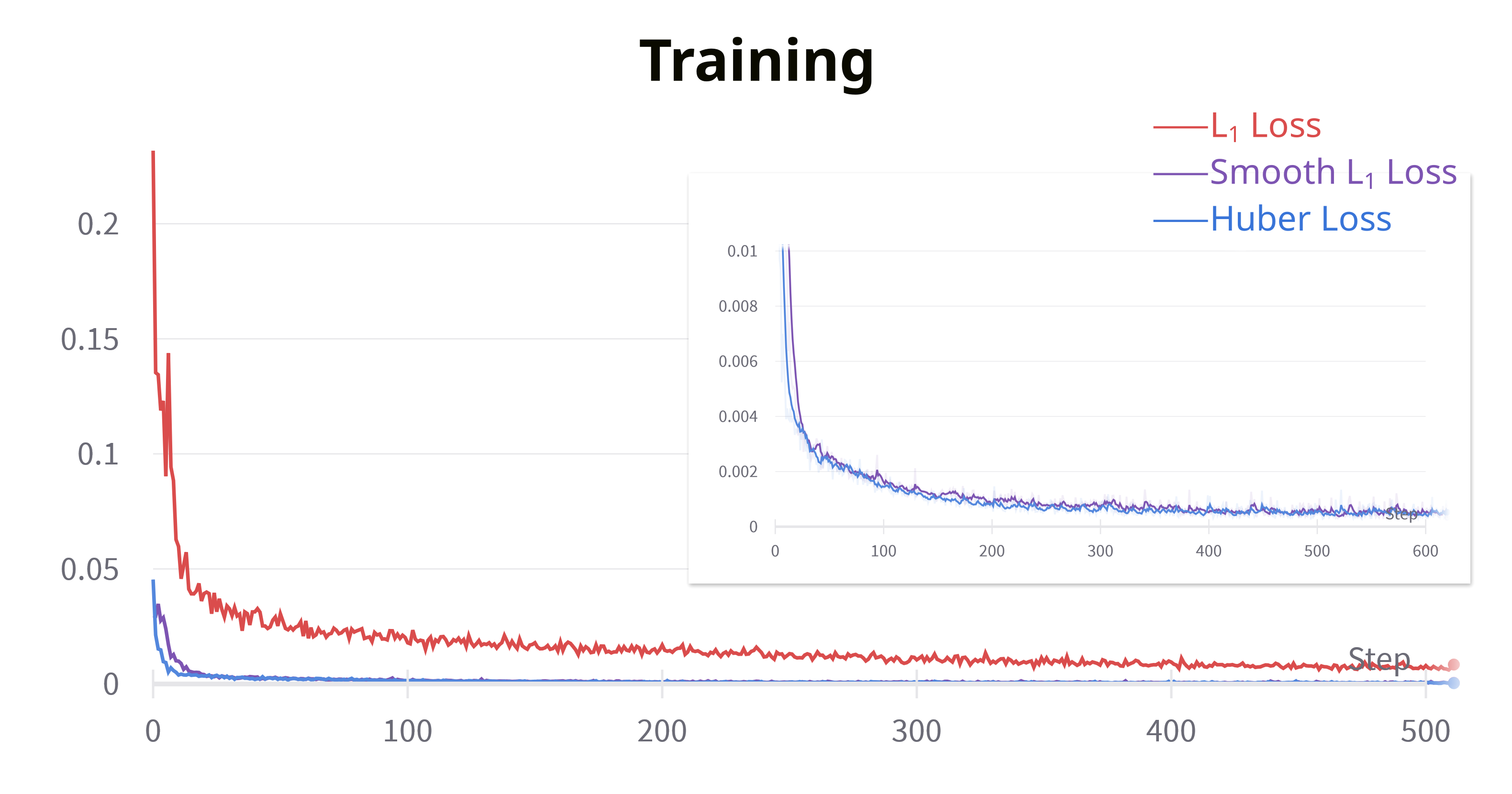}%
        \label{fig: samllLosses}
    }
    \hfil
    \subfloat[]{\includegraphics[height=5cm]{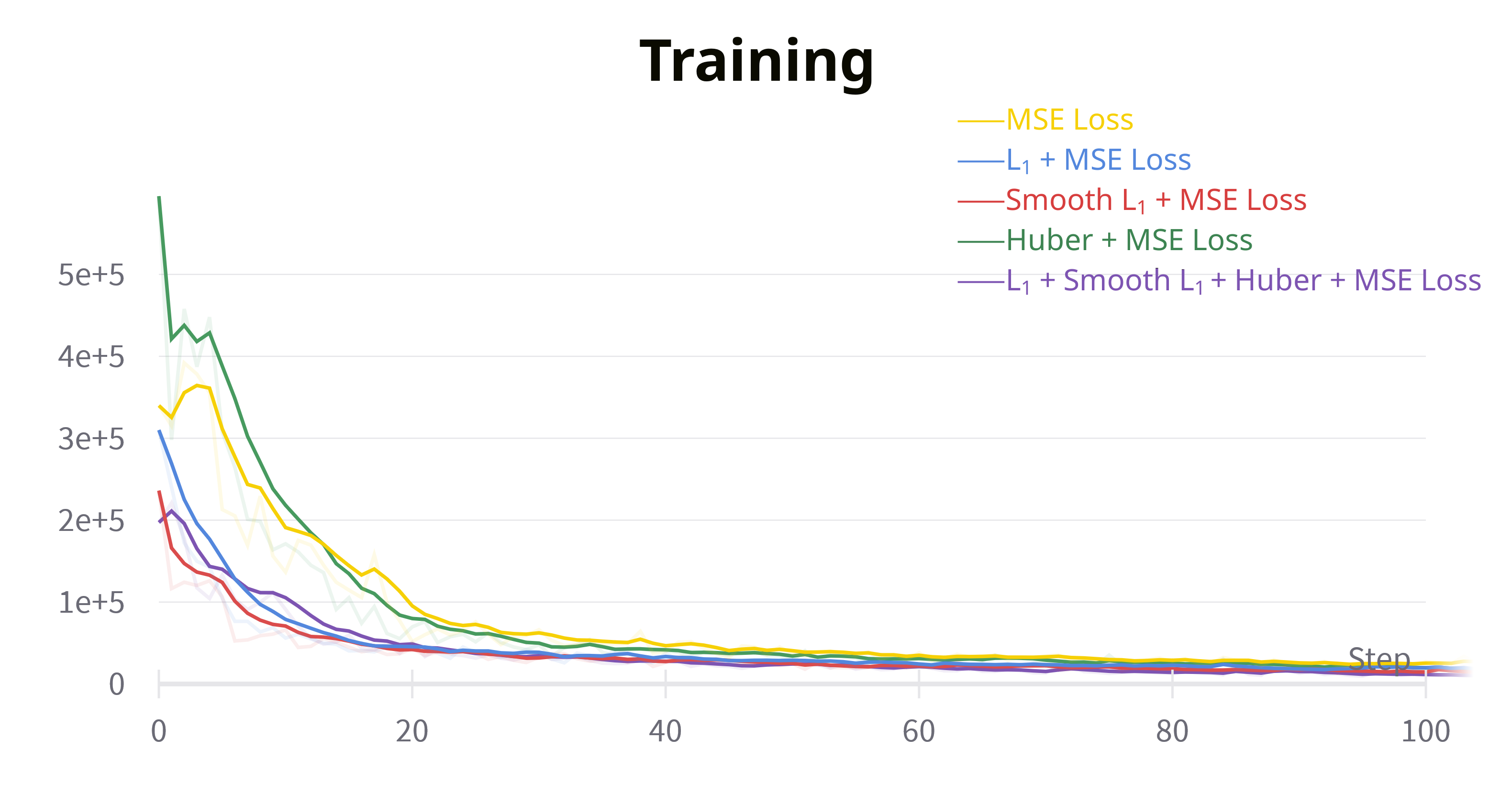}%
        \label{fig: biglosses}
    }
    \hfil
    \caption{Loss functions convergence comparison: (a) Loss of $L_1$ and its variants. (b) Loss of MSE loss and other combined losses.}
    \label{fig: losses}
\end{figure*}

To find the optimal loss function, we evaluate the convergence speed of different loss functions.
The loss functions we tested include $L_1$ loss, Huber loss, Smooth $L_1$ loss, MSE loss and several combinations of aforementioned loss functions.
The result is shown in Fig. \ref{fig: losses}.

In order to better visualize the differences in convergence speed between the losses, we present them in separated subplots. 
As shown in Fig. \ref{fig: samllLosses}, the $L_1$ loss and its variants (Huber loss and Smooth $L_1$ loss) are displayed on one subplot,
while the MSE loss-related losses are presented on another subplot in Fig. \ref{fig: biglosses}. 

We observed that implementing MSE loss results in faster convergence, allowing the model to reach convergence in under 100 steps, as shown in Fig. \ref{fig: biglosses}.
Meanwhile, as depicted in Fig. \ref{fig: samllLosses}, the loss functions based on $L_1$ loss achieve a much slower convergence after approximately 500 to 600 steps. 
Although Huber loss and Smooth $L_1$ loss seem to have a quicker rate of convergence, closer examination in Fig. \ref{fig: samllLosses} reveals that they both take around 500 steps to converge, which is similar to the standard $L_1$ loss.

We also noted from Fig. \ref{fig: biglosses} that using a combination of MSE loss and different $L_1$ based losses doesn't significantly affect the rate of convergence, likely because the difference in scale between the MSE loss and L1 loss and its variants causes MSE loss to remain the primary determinant of convergence speed.

Our study also conducted an evaluation of the costumed U-Net trained using various loss functions.
Our findings in Tables \ref{tab:loss_functions_results} and \ref{tab:loss_functions_results_recon} revealed that there was minimal difference between the performances of these models, with the largest discrepancies in Dice, IoU, PA, SSIM and PSNR\_HVS\_M amounting to $0.023$, $0.019$, $0.003$, $0.011$ and $4.031$ respectively.
These outcomes suggest that the selection of alternative loss functions has little influence on the overall performance of the model. 
As such, we decided not to employ the MSE loss function in subsequent experiments and instead continued to utilize the $L_1$ loss.

\begin{table*}[ht]
    \centering
    \caption{Segmentation Precision on Body Marker Annotation for the Costumed U-Net Trained with Different Loss Functions (Average + Var)}
    \begin{tabular}{|c c c c|}
        \hline
        Loss Function  & Dice            & IoU             & PA               \\ [0.5ex]
        \hline
        $L_1$             & $0.712\pm0.053$ & $0.596\pm0.058$ & $0.993\pm 0.007$ \\
        \hline
        Huber          & $0.708\pm0.05$  & $0.592\pm0.005$ & $0.993\pm 0.005$ \\
        \hline
        Smooth $L_1$      & $0.717\pm0.05$  & $0.599\pm0.055$ & $0.993\pm 0.005$ \\
        \hline
        L2             & $0.716\pm0.053$ & $0.599\pm0.056$ & $0.993\pm 0.005$ \\
        \hline
        $L_1$ + L2        & $0.712\pm0.053$ & $0.596\pm0.057$ & $0.993\pm 0.005$ \\
        \hline
        Huber + L2     & $0.713\pm0.052$ & $0.596\pm0.057$ & $0.993\pm 0.005$ \\
        \hline
        Smooth $L_1$ + L2 & $0.692\pm0.068$ & $0.580\pm0.066$ & $0.990\pm 0.005$ \\
        \hline
        All Loss Sum   & $0.715\pm0.052$ & $0.598\pm0.057$ & $0.993\pm 0.005$ \\
        \hline
    \end{tabular}
    \label{tab:loss_functions_results}
\end{table*}

\begin{table*}[ht]
    \centering
    \caption{Reconstruction Similarity on Body Marker Annotation for the Costumed U-Net Trained with Different Loss Functions (Average + Var)}
    \begin{tabular}{|c c c |}
        \hline
        Loss Function  & SSIM            & PSNR\_HVS\_M    \\ [0.5ex]
        \hline
        $L_1$             & $0.967\pm0.000$ & $41.628\pm41.775$ \\
        \hline
        Huber          & $0.968\pm0.000$  & $38.110\pm91.416$ \\
        \hline
        Smooth $L_1$      & $0.967\pm0.000$  & $41.737\pm38.719$ \\
        \hline
        L2             & $0.966\pm0.000$ & $40.982\pm37.608$ \\
        \hline
        $L_1$ + L2        & $0.966\pm0.000$ & $38.689\pm46.355$ \\
        \hline
        Huber + L2     & $0.977\pm0.000$ & $42.141\pm53.215$ \\
        \hline
        Smooth $L_1$ + L2 & $0.968\pm0.000$ & $39.186\pm47.249$ \\
        \hline
        All Loss Sum   & $0.968\pm0.000$ & $40.443\pm57.084$ \\
        \hline
    \end{tabular}
    \label{tab:loss_functions_results_recon}
\end{table*}

\subsection{Noise2Noise with Other Annotations}

The improvement observed in the costumed U-Net trained using the Noise2Noise scheme is also apparent in other annotation datasets, as shown in \Cref{tab:measure_anchor_seg,tab:measure_anchor_restore,tab:color_rectangle_seg,tab:color_rectangle_restore}. 
In the provided tables, the costumed U-Net has been trained using other two annotation datasets along with two different training schemes. The outcomes show a substantial enhancement in comparison to the Noise2Clean models, as there is approximately a half unit gain observed in both Dice and IoU metrics, an increase of around $0.01$ in SSIM, and a rise of $5$ units in PSNR\_HVS\_M for both types of annotations.

\begin{table*}[ht]
    \centering
    \caption{Segmentation Precision on Radial Line Annotation (Average + Var)}
    \begin{tabular}{|c| c c c c|}
        \hline
        Method                            & Training Mode & Dice                   & IoU                    & PA                                  \\
        \hline
        \multirow{2}{*}{Costumed U-Net  } & N2C           & $0.226 \pm 0.013$      & $0.132 \pm 0.006$      & $0.992 \pm 6.63 \times e^{-6}$      \\
        \cline{2-5}
                                          & \textbf{N2N}  & $\bm{0.747 \pm 0.004}$ & $\bm{0.639 \pm 0.059}$ & $\bm{0.999 \pm 3.54 \times e^{-6}}$ \\
        \hline
    \end{tabular}
    \label{tab:measure_anchor_seg}
\end{table*}

\begin{table*}[ht]
    \centering
    \caption{Reconstruction Similarity on Radial Line Annotation (Average + Var)}
    \begin{tabular}{|c| c c c |}
        \hline
        Method                            & Training Mode & SSIM                   & PSNR\_HVS\_M            \\[0.5ex]
        \hline
        \multirow{2}{*}{Costumed U-Net  } & N2C           & $0.932 \pm 0.000$      & $21.660\pm 5.391$       \\
        \cline{2-4}
                                          & \textbf{N2N}  & $\bm{0.942 \pm 0.000}$ & $\bm{26.376 \pm 0.681}$ \\
        \hline
    \end{tabular}
    \label{tab:measure_anchor_restore}
\end{table*}

\begin{table*}[ht]
    \centering
    \caption{Segmentation Precision on Vascular Flow Annotation (Average + Var)}
    \begin{tabular}{|c |c c c c|}
        \hline
        Method                            & Training Mode & Dice                   & IoU                    & PA                                    \\
        \hline
        \multirow{2}{*}{Costumed U-Net  } & N2C           & $0.243 \pm 0.028$      & $0.149 \pm 0.013$      & $0.989 \pm 1.115$                     \\
        \cline{2-5}
                                          & \textbf{N2N}  & $\bm{0.728 \pm 0.031}$ & $\bm{0.599 \pm 0.039}$ & $\bm{0.998 \pm 1.423 \times e^{-05}}$ \\
        \hline
    \end{tabular}
    \label{tab:color_rectangle_seg}
\end{table*}

\begin{table*}[ht]
    \centering
    \caption{Reconstruction Similarity on Vascular Flow Annotation (Average + Var)}
    \begin{tabular}{|c| c c c |}
        \hline
        Method                            & Training Mode & SSIM                                 & PSNR\_HVS\_M            \\[0.5ex]
        \hline
        \multirow{2}{*}{Costumed U-Net  } & N2C           & $0.938 \pm 0.000$                    & $21.584\pm 5.384$       \\
        \cline{2-4}
                                          & \textbf{N2N}  & $\bm{0.948 \pm 4.853 \times e^{-5}}$ & $\bm{26.717 \pm 0.511}$ \\
        \hline
    \end{tabular}
    \label{tab:color_rectangle_restore}
\end{table*}

\subsection{Qualitative Results}
In this section, we present denoised images from models trained under different schemes to further support our claim.

\begin{figure*}[!t]
    \centering
    \subfloat[]{\includegraphics[height=5cm]{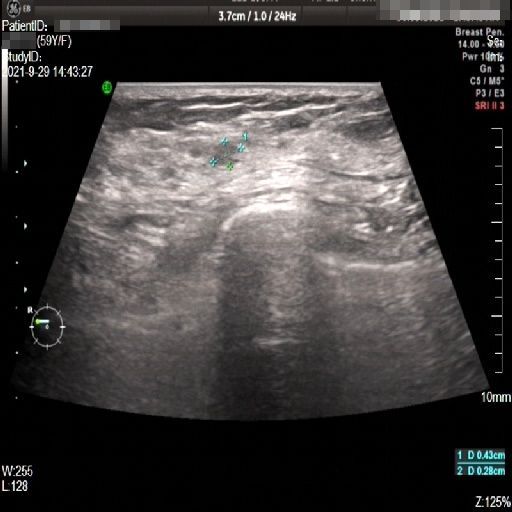}
        \label{fig: bn2c_0i}}
    \hfil
    \subfloat[]{\includegraphics[height=5cm]{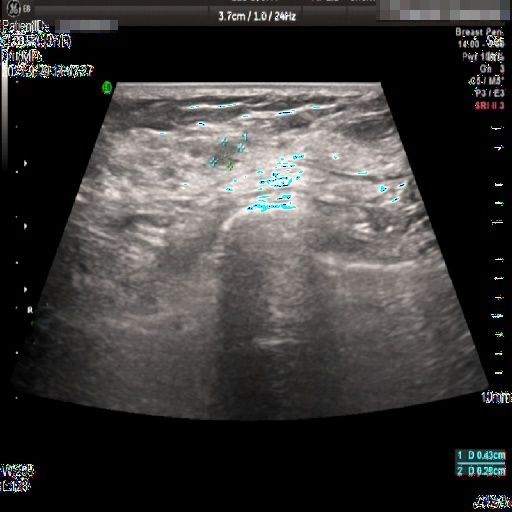}
        \label{fig: bn2c_0O}}
    \hfil
    \subfloat[]{\includegraphics[height=5cm]{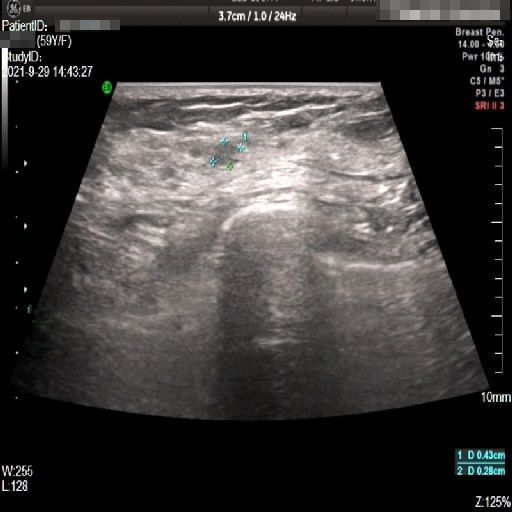}
        \label{fig: bn2n_0O}}
    \caption{Body marker annotation: (a) input image, (b) output from N2C model, (c) output from N2N model.}
    \label{fig: quali_body}
\end{figure*}

\begin{figure*}[!t]
    \centering
    \subfloat[]{\includegraphics[height=5cm]{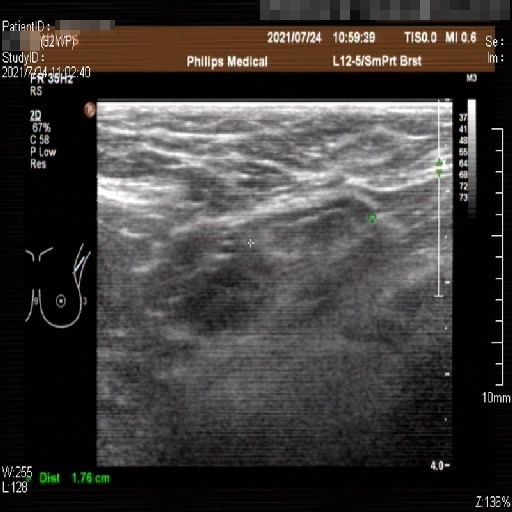}
        \label{fig: an2c_0i}}
    \hfil
    \subfloat[]{\includegraphics[height=5cm]{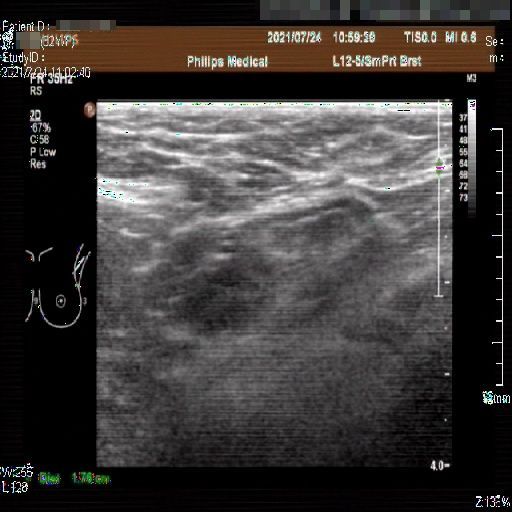}
        \label{fig: an2c_0o}}
    \hfil
    \subfloat[]{\includegraphics[height=5cm]{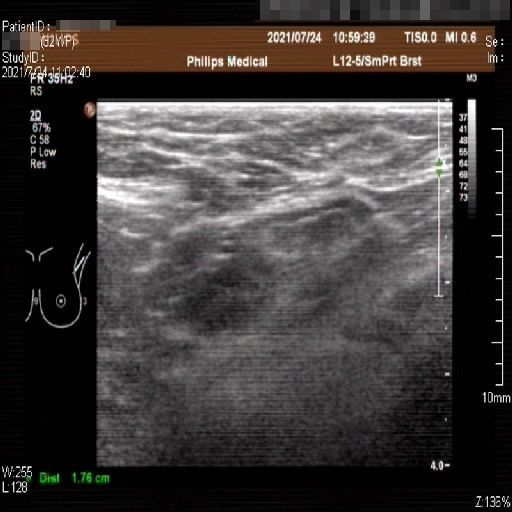}
        \label{fig: an2n_0o}}
    \caption{Radial line annotation: (a) input image, (b) output from N2C model, (c) output from N2N model.}
    \label{fig: quali_anchor}
\end{figure*}

\begin{figure*}[!t]
    \centering
    \subfloat[]{\includegraphics[height=5cm]{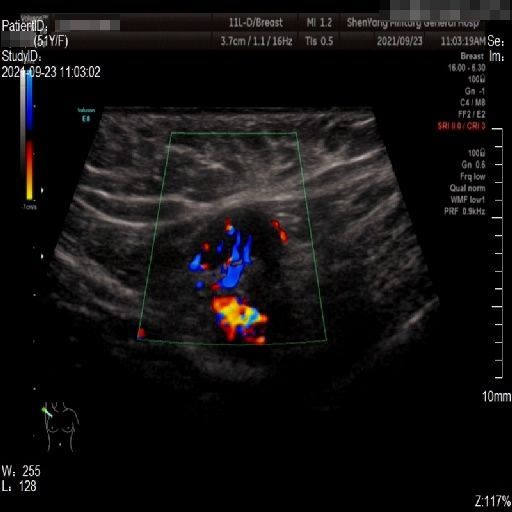}
        \label{fig: vn2c_19i}}
    \hfil
    \subfloat[]{\includegraphics[height=5cm]{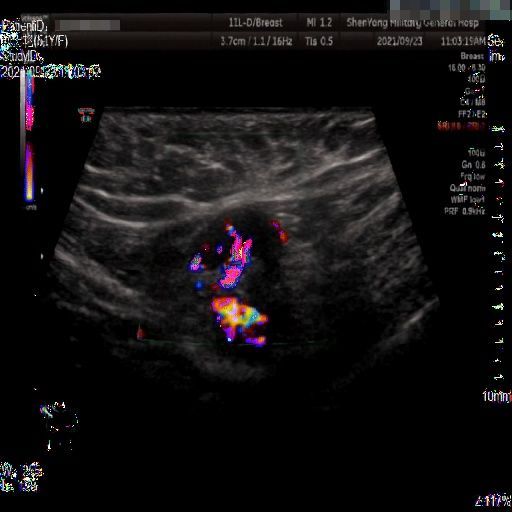}
        \label{fig: vn2c_19o}}
    \hfil
    \subfloat[]{\includegraphics[height=5cm]{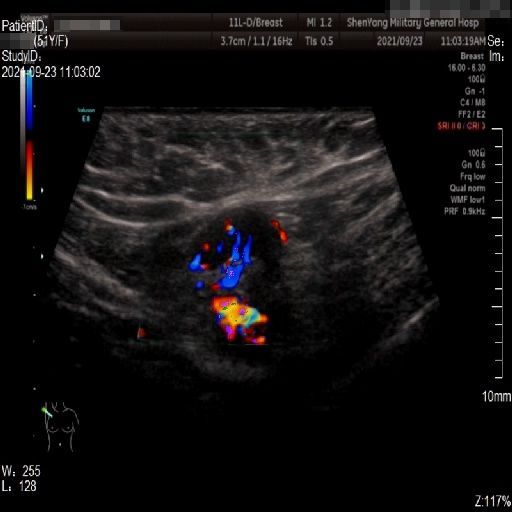}
        \label{fig: vn2n_19o}}
    \caption{Vascular flow annotation: (a) input image, (b) output from N2C model, (c) output from N2N model.}
    \label{fig: quali_vascularc}
\end{figure*}

As can be seen in Figs. \ref{fig: quali_body}, \ref{fig: quali_anchor} and \ref{fig: quali_vascularc},
the output from the Noise2Clean model contains obvious artifacts, whereas models trained using the Noise2Noise scheme do not suffer from this problem.

It is also worth noting that in the output images from Noise2Clean models, information in the edge area is compromised.
In contrast, the Noise2Noise models preserve this information well.
The evidence implies that models trained with the Noise2Noise scheme possess superior capabilities in identifying and distinguishing noise.


\section{Discussion}\label{sec: discuss}
This study proposed a self-supervised data generation and training approach to build a large and diverse datasets starting from a small dataset with only few clean images.
We find that the costumed U-Net trained with the Noise2Noise scheme outperformed other models in terms of segmentation precision and reconstruction similarity in the annotation removal task. 
The benefits of Noise2Noise training were observed across most model structures tested, and the models trained using this scheme produced fewer artifacts. 

Our study has some limitations:
Firstly, we used separate parameter sets for the segmentation task of different annotations. 
However, with the recent advancement of deep learning theories, it is now possible to use a single parameter set for the segmentation of all annotations presented in the image.
Additionally, there is potential for further research in the area of language-guided segmentation models, which would provide a more precise and flexible interface for medical professionals. 
We find building a model that incorporates these innovations intriguing.

We also noted that our model was trained in a self-supervised manner, meaning it has potentially gained a strong understanding of the structural features of ultrasonic images. 
This understanding is beneficial for downstream models such as object detection model.
Different ways of fine-tuning, like Low-Rank Adaptation (LoRA), adapter layers, etc. should be explored to find the optimal method to effectively transfer this understanding.
We plan to address these issues in future studies.

\nocite{*}

\bibliography{references}

\begin{thebibliography}{28}
\providecommand{\natexlab}[1]{#1}
\providecommand{\url}[1]{#1}
\csname url@samestyle\endcsname
\providecommand{\newblock}{\relax}
\providecommand{\bibinfo}[2]{#2}
\providecommand{\BIBentrySTDinterwordspacing}{\spaceskip=0pt\relax}
\providecommand{\BIBentryALTinterwordstretchfactor}{4}
\providecommand{\BIBentryALTinterwordspacing}{\spaceskip=\fontdimen2\font plus
\BIBentryALTinterwordstretchfactor\fontdimen3\font minus \fontdimen4\font\relax}
\providecommand{\BIBforeignlanguage}[2]{{%
\expandafter\ifx\csname l@#1\endcsname\relax
\typeout{** WARNING: IEEEtranN.bst: No hyphenation pattern has been}%
\typeout{** loaded for the language `#1'. Using the pattern for}%
\typeout{** the default language instead.}%
\else
\language=\csname l@#1\endcsname
\fi
#2}}
\providecommand{\BIBdecl}{\relax}
\BIBdecl

\bibitem[Kulshrestha and Singh(2016)]{kulshrestha2016inter}
A.~Kulshrestha and J.~Singh, ``Inter-hospital and intra-hospital patient transfer: Recent concepts,'' \emph{Indian journal of anaesthesia}, vol.~60, no.~7, p. 451, 2016.

\bibitem[Jackson and Chenal(2017)]{jackson2017ultrasonic}
P.~Jackson and C.~Chenal, ``Ultrasonic imaging system with body marker annotations,'' 7 2017, uS Patent 9,713,458.

\bibitem[Li \textit{et~al.}(2018)Li, Xiong, An, and Wang]{li2018pyramid}
H.~Li, P.~Xiong, J.~An, and L.~Wang, ``Pyramid attention network for semantic segmentation,'' \emph{arXiv preprint arXiv:1805.10180}, 2018.

\bibitem[Huang \textit{et~al.}(2017)Huang, Xia, Wu, Li, Wang, Song, and Kuo]{huang2017semantic}
Q.~Huang, C.~Xia, C.~Wu, S.~Li, Y.~Wang, Y.~Song, and C.-C.~J. Kuo, ``Semantic segmentation with reverse attention,'' \emph{arXiv preprint arXiv:1707.06426}, 2017.

\bibitem[Goodfellow \textit{et~al.}(2020)Goodfellow, Pouget-Abadie, Mirza, Xu, Warde-Farley, Ozair, Courville, and Bengio]{goodfellow2020generative}
I.~Goodfellow, J.~Pouget-Abadie, M.~Mirza, B.~Xu, D.~Warde-Farley, S.~Ozair, A.~Courville, and Y.~Bengio, ``Generative adversarial networks,'' \emph{Communications of the ACM}, vol.~63, no.~11, pp. 139--144, 2020.

\bibitem[Liu \textit{et~al.}(2016)Liu, Anguelov, Erhan, Szegedy, Reed, Fu, and Berg]{liu2016ssd}
W.~Liu, D.~Anguelov, D.~Erhan, C.~Szegedy, S.~Reed, C.-Y. Fu, and A.~C. Berg, ``Ssd: Single shot multibox detector,'' in \emph{Computer Vision--ECCV 2016: 14th European Conference, Amsterdam, The Netherlands, October 11--14, 2016, Proceedings, Part I 14}.\hskip 1em plus 0.5em minus 0.4em\relax Springer, 2016, pp. 21--37.

\bibitem[Redmon \textit{et~al.}(2016)Redmon, Divvala, Girshick, and Farhadi]{redmon2016you}
J.~Redmon, S.~Divvala, R.~Girshick, and A.~Farhadi, ``You only look once: Unified, real-time object detection,'' in \emph{Proceedings of the IEEE conference on computer vision and pattern recognition}, 2016, pp. 779--788.

\bibitem[Lehtinen \textit{et~al.}(2018)Lehtinen, Munkberg, Hasselgren, Laine, Karras, Aittala, and Aila]{lehtinen2018Noise2Noise}
J.~Lehtinen, J.~Munkberg, J.~Hasselgren, S.~Laine, T.~Karras, M.~Aittala, and T.~Aila, ``Noise2noise: Learning image restoration without clean data,'' \emph{arXiv preprint arXiv:1803.04189}, 2018.

\bibitem[Kashyap \textit{et~al.}(2021)Kashyap, Tambwekar, Manohara, and Natarajan]{kashyap2021speech}
M.~M. Kashyap, A.~Tambwekar, K.~Manohara, and S.~Natarajan, ``Speech denoising without clean training data: a noise2noise approach,'' \emph{arXiv preprint arXiv:2104.03838}, 2021.

\bibitem[Long \textit{et~al.}(2015)Long, Shelhamer, and Darrell]{long2015fully}
J.~Long, E.~Shelhamer, and T.~Darrell, ``Fully convolutional networks for semantic segmentation,'' in \emph{Proceedings of the IEEE conference on computer vision and pattern recognition}, 2015, pp. 3431--3440.

\bibitem[Minaee \textit{et~al.}(2021)Minaee, Boykov, Porikli, Plaza, Kehtarnavaz, and Terzopoulos]{minaee2021image}
S.~Minaee, Y.~Y. Boykov, F.~Porikli, A.~J. Plaza, N.~Kehtarnavaz, and D.~Terzopoulos, ``Image segmentation using deep learning: A survey,'' \emph{IEEE transactions on pattern analysis and machine intelligence}, 2021.

\bibitem[Chen \textit{et~al.}(2017)Chen, Papandreou, Schroff, and Adam]{chen2017rethinking}
L.-C. Chen, G.~Papandreou, F.~Schroff, and H.~Adam, ``Rethinking atrous convolution for semantic image segmentation,'' \emph{arXiv preprint arXiv:1706.05587}, 2017.

\bibitem[Chaurasia and Culurciello(2017)]{chaurasia2017linknet}
A.~Chaurasia and E.~Culurciello, ``Linknet: Exploiting encoder representations for efficient semantic segmentation,'' in \emph{2017 IEEE visual communications and image processing (VCIP)}.\hskip 1em plus 0.5em minus 0.4em\relax IEEE, 2017, pp. 1--4.

\bibitem[Li \textit{et~al.}(2021)Li, Zheng, Zhang, Duan, Su, Wang, and Atkinson]{li2021multiattention}
R.~Li, S.~Zheng, C.~Zhang, C.~Duan, J.~Su, L.~Wang, and P.~M. Atkinson, ``Multiattention network for semantic segmentation of fine-resolution remote sensing images,'' \emph{IEEE Transactions on Geoscience and Remote Sensing}, vol.~60, pp. 1--13, 2021.

\bibitem[Iakubovskii(2019)]{Iakubovskii:2019}
P.~Iakubovskii, ``Segmentation models pytorch,'' \url{https://github.com/qubvel/segmentation_models.pytorch}, 2019.

\bibitem[Ronneberger \textit{et~al.}(2015)Ronneberger, Fischer, and Brox]{ronneberger2015u}
O.~Ronneberger, P.~Fischer, and T.~Brox, ``U-net: Convolutional networks for biomedical image segmentation,'' in \emph{Medical Image Computing and Computer-Assisted Intervention--MICCAI 2015: 18th International Conference, Munich, Germany, October 5-9, 2015, Proceedings, Part III 18}.\hskip 1em plus 0.5em minus 0.4em\relax Springer, 2015, pp. 234--241.

\bibitem[Chlap \textit{et~al.}(2021)Chlap, Min, Vandenberg, Dowling, Holloway, and Haworth]{chlap2021review}
P.~Chlap, H.~Min, N.~Vandenberg, J.~Dowling, L.~Holloway, and A.~Haworth, ``A review of medical image data augmentation techniques for deep learning applications,'' \emph{Journal of Medical Imaging and Radiation Oncology}, vol.~65, no.~5, pp. 545--563, 2021.

\bibitem[Zhou \textit{et~al.}(2018)Zhou, Rahman~Siddiquee, Tajbakhsh, and Liang]{zhou2018unet++}
Z.~Zhou, M.~M. Rahman~Siddiquee, N.~Tajbakhsh, and J.~Liang, ``Unet++: A nested u-net architecture for medical image segmentation,'' in \emph{Deep Learning in Medical Image Analysis and Multimodal Learning for Clinical Decision Support: 4th International Workshop, DLMIA 2018, and 8th International Workshop, ML-CDS 2018, Held in Conjunction with MICCAI 2018, Granada, Spain, September 20, 2018, Proceedings 4}.\hskip 1em plus 0.5em minus 0.4em\relax Springer, 2018, pp. 3--11.

\bibitem[Ibtehaz and Rahman(2020)]{ibtehaz2020multiresunet}
N.~Ibtehaz and M.~S. Rahman, ``Multiresunet: Rethinking the u-net architecture for multimodal biomedical image segmentation,'' \emph{Neural networks}, vol. 121, pp. 74--87, 2020.

\bibitem[Ponomarenko \textit{et~al.}(2007)Ponomarenko, Silvestri, Egiazarian, Carli, Astola, and Lukin]{ponomarenko2007betweenpsnrhvsm}
N.~Ponomarenko, F.~Silvestri, K.~Egiazarian, M.~Carli, J.~Astola, and V.~Lukin, ``On between-coefficient contrast masking of dct basis functions,'' in \emph{Proceedings of the third international workshop on video processing and quality metrics}, vol.~4.\hskip 1em plus 0.5em minus 0.4em\relax Scottsdale USA, 2007.

\bibitem[Tieleman \textit{et~al.}(2012)Tieleman, Hinton, \textit{et~al.}]{tieleman2012lecture}
T.~Tieleman, G.~Hinton \emph{\textit{et~al.}}, ``Lecture 6.5-rmsprop: Divide the gradient by a running average of its recent magnitude,'' \emph{COURSERA: Neural networks for machine learning}, vol.~4, no.~2, pp. 26--31, 2012.

\bibitem[Foundation()]{rmsproptorch}
\BIBentryALTinterwordspacing
T.~P. Foundation, ``Rmsprop.'' [Online]. Available: \url{https://pytorch.org/docs/stable/generated/torch.optim.RMSprop.html#torch.optim.RMSprop}
\BIBentrySTDinterwordspacing

\bibitem[Calvarons(2021)]{calvarons2021improved}
A.~F. Calvarons, ``Improved noise2noise denoising with limited data,'' in \emph{Proceedings of the IEEE/CVF Conference on Computer Vision and Pattern Recognition}, 2021, pp. 796--805.

\bibitem[Egiazarian \textit{et~al.}(2006)Egiazarian, Astola, Ponomarenko, Lukin, Battisti, and Carli]{egiazarian2006new}
K.~Egiazarian, J.~Astola, N.~Ponomarenko, V.~Lukin, F.~Battisti, and M.~Carli, ``New full-reference quality metrics based on hvs,'' in \emph{Proceedings of the second international workshop on video processing and quality metrics}, vol.~4, 2006.

\bibitem[Hasan \textit{et~al.}(2020)Hasan, Mohebbian, Wahid, and Babyn]{hasan2020hybrid}
A.~M. Hasan, M.~R. Mohebbian, K.~A. Wahid, and P.~Babyn, ``Hybrid-collaborative noise2noise denoiser for low-dose ct images,'' \emph{IEEE Transactions on Radiation and Plasma Medical Sciences}, vol.~5, no.~2, pp. 235--244, 2020.

\bibitem[Hussain \textit{et~al.}(2017)Hussain, Gimenez, Yi, and Rubin]{hussain2017differential}
Z.~Hussain, F.~Gimenez, D.~Yi, and D.~Rubin, ``Differential data augmentation techniques for medical imaging classification tasks,'' in \emph{AMIA annual symposium proceedings}, vol. 2017.\hskip 1em plus 0.5em minus 0.4em\relax American Medical Informatics Association, 2017, p. 979.

\bibitem[Qiu \textit{et~al.}(2021)Qiu, Zeng, Meng, Jiang, You, Geng, Li, Hu, Huang, Zhou, \textit{et~al.}]{qiu2021comparative}
B.~Qiu, S.~Zeng, X.~Meng, Z.~Jiang, Y.~You, M.~Geng, Z.~Li, Y.~Hu, Z.~Huang, C.~Zhou \emph{\textit{et~al.}}, ``Comparative study of deep neural networks with unsupervised noise2noise strategy for noise reduction of optical coherence tomography images,'' \emph{Journal of Biophotonics}, vol.~14, no.~11, p. e202100151, 2021.

\bibitem[Wang \textit{et~al.}(2004)Wang, Bovik, Sheikh, and Simoncelli]{wang2004image}
Z.~Wang, A.~C. Bovik, H.~R. Sheikh, and E.~P. Simoncelli, ``Image quality assessment: from error visibility to structural similarity,'' \emph{IEEE transactions on image processing}, vol.~13, no.~4, pp. 600--612, 2004.

\end{thebibliography}

\end{document}